\documentstyle[twoside,fleqn,espcrc2,epsf]{article}
\newcommand{\AmS}{{\protect\the\textfont2
  A\kern-.1667em\lower.5ex\hbox{M}\kern-.125emS}}
\newcommand{\beq}{\begin{equation}}
\newcommand{\eeq}{\end{equation}}
\newcommand{\bea}{\begin{eqnarray}}
\newcommand{\eea}{\end{eqnarray}}

\newcommand{\U}{\hat{U}}

\title{$Z(2)$ vortices and the string tension in $SU(2)$ 
        gauge theory\thanks{Talk 
presented by E. T. Tomboulis}}

\author{Tam\'as G. Kov\'acs\address{Department of Physics, 
        University of Colorado\\ Boulder, CO 80309-390, USA}
        \thanks{Research partially supported by NSF PHY-9023257, 
         DE-FG02-92ER-40672} 
        and 
        E. T. Tomboulis\address{Department of Physics, 
        UCLA\\ 
        Los Angeles, CA 90095-1547, USA}
        \thanks{Research partially supported by NSF 
         PHY-95310223.}}
       
\begin{document}

\begin{abstract}
We exhibit the appropriate variables allowing the 
plus-minus ($Z(2)$) `reduction' of the Wilson loop 
operator which provides a direct measure of the thin, 
thick and `mixed' $Z(2)$ topological, gauge-invariant 
vortices in SU(2) LGT. Simulations with the Wilson action, 
as well as a perfect action smoothing 
procedure, show the string tension to be reproduced from 
the contributions of these excitations. 
\end{abstract}

\maketitle

We report on recent work on the crucial role of  
vortices characterized by $Z(N)$ flux for maintaining 
confinement at weak coupling in $SU(N)$ gauge theory.   
The present work is a 
continuation of our on-going project over the last 
several years \cite{T}. 

We begin by distinguishing between various types of vortices. 
Recall that in the continuum formulation  
there is no local distinction between pure $SU(N)$ and 
$SU(N)/Z(N)$ gauge theories, but in the lattice formulation 
there is. Now for continuum 
$SU(N)/Z(N)$ fields, vortices are topologically classified by 
$\pi_1(SU(N)/Z(N))=Z(N)$. A vortex forms a closed 2-dim 
structures in $d=4$. Topologically, 
it is also possible to have Dirac monopoles, also classified 
by $\pi_1(SU(N)/Z(N))$. The Dirac sheet of such a monopole loop 
($d=4$) may be described as defining a `punctured' vortex. 
On the lattice the $SU(N)$ and $SU(N)/Z(N)$ 
theories differ by $Z(N)$ degrees of freedom. Exciting these 
$Z(N)$ degrees of freedom gives rise to `thin' $Z(N)$ vortices. 
They are very efficient at disordering at small $\beta$; but 
are directly suppressed by the $SU(N)$ plaquette action and 
become unimportant at large $\beta$. This reflects the fact 
that the distinction between $SU(N)$ and $SU(N)/Z(N)$
LGT must disappear as the continuum is approached.  

Failure to properly distinguish between thick and thin, 
i.e. between the (lattice analogs of the) 
$\pi_1(SU(N)/Z(N))=Z(N)$ excitations arising from the  
$SU(N)/Z(N)$ part of the $SU(N)$ gauge group versus the 
$Z(N)$ excitations of the $Z(N)$ part of the group, has 
caused confusion in the lattice literature. A clean separation 
can be achieved by introducing new separate $SU(N)/Z(N)$ 
and $Z(N)$ variables. We treat explicitly the $N=2$ case which 
is the actual case of our numerical simulations; the extension 
to general $N$ is straightforward. The original $SU(2)$ LGT 
is given in terms of the bond variables $U_b \in SU(2)$ residing 
on bonds $b$. The new variables are coset bond variables $\U_b 
\in SU(2)/Z(2) \sim SO(3)$, and $Z(2)$ variables $\sigma_p 
\in \{\pm1\}$ residing on plaquettes $p$. The $SU(2)$ theory, i.e. 
the partition function and all correlations, can be rewritten in 
terms of the variables $\U_b$ and $\sigma_p$ \cite{T}. This 
rewritting is exact and gauge invariant and 
expresses the partition function as a coupled $SO(3)-Z(2)$ 
system over the Haar invariant measures of the two groups. 
The $SU(2)$ plaquette action in the new variables 
becomes the product of a non-negative function of the $\U_b$'s 
times $\sigma_p$. Thus the sign of the $SU(2)$ action 
on a plaquette $p$ is now simply given by the 
variable $\sigma_p$. The integration measure in the new 
variables contains a constraint that requires 
that the product of the $\sigma_p$'s over the faces of a cube 
equals $-1$ if the cube is the site of a Dirac monopole in the 
$\{\U\}$ configuration. 
\begin{figure}[tb]
\begin{minipage}{7.5cm}
\epsfxsize=7cm \epsfysize=7cm
\epsfbox{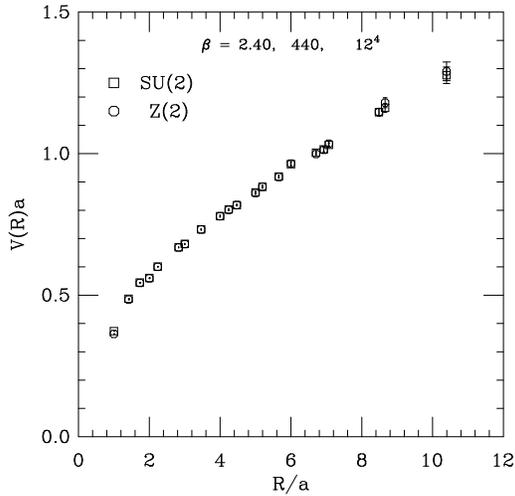}
\end{minipage}
\vspace{-0.7cm}
\caption{\label{fig:pot12_b2.4}The heavy quark potential 
measured on an ensemble of
440 12$^4$ configurations with Wilson action at $\beta=2.4$.}
\end{figure}

The new variables allow an immediate identification of 
the various possible excitations. 
Thin vortices are created by the excitation of the $\sigma_p$ 
variables. They are necessarily localized to one lattice 
spacing thickness, and incur a direct action cost 
proportional to their vortex sheet area. At large $\beta$ 
then, long thin vortices are suppressed, 
and only short thin vortices survive. `Thick' vortices 
are vortices in the $\{\U_b\}$ configurations. There 
is no negative plaquette action supression associated 
with them, so that, by being sufficiently spread out, 
these vortex configurations can cost locally 
very little action even at large $\beta$; 
while, by the very gradual variation of 
the bond variables, can  disorder the system over long 
scales. `Punctured' thick vortices can also exist at large 
$\beta$ if the `hole', whose boundary is a Dirac monopole 
current loop, is small (of the order of one lattice 
spacing for sufficiently large $\beta$). This is 
because the hole has to be covered 
by an open thin vortex sheet to satisfy the above constraint 
on the cubes occupied by the monopoles forming its 
boundary. The result is a `hybrid' vortex. Hybrid vortices, 
whose presence was explored in \cite{T}, act then in 
the same way as pure thick vortices to disorder over long 
scales.     
\begin{figure}[tb]
\begin{minipage}{7.5cm}
\epsfxsize=7cm \epsfysize=7cm
\epsfbox{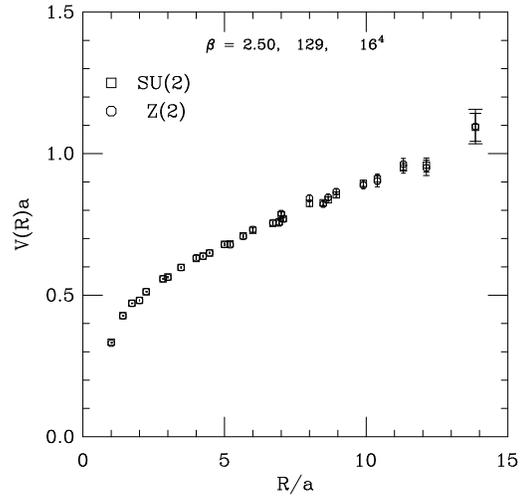}
\end{minipage}
\vspace{-0.7cm}
\caption{\label{fig:pot16_b2.5}
The heavy quark potential measured on an ensemble of
129 16$^4$ configurations with Wilson action at $\beta=2.5$.}
\end{figure}

\begin{figure}[!tb]
\begin{minipage}{7.5cm}
\epsfxsize=7cm \epsfysize=7cm
\epsfbox{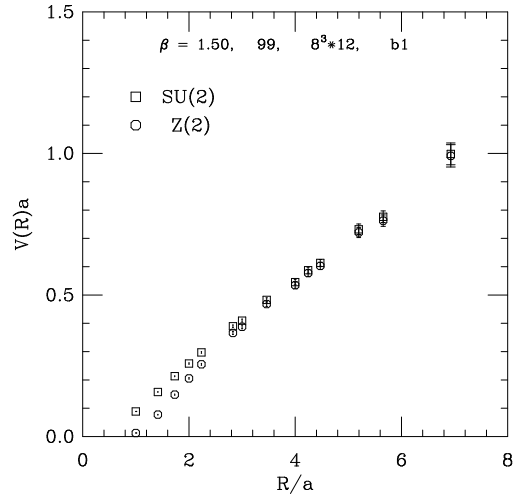}
\end{minipage}
\vspace{-0.7cm}
\caption{\label{fig:pot812_b1.5_b1}
The heavy quark potential measured after 1 smoothing step 
on an ensemble of 99 $8^3*12$ configurations generated 
with a fixed point action at $\beta=1.5$ which corresponds to 
very nearly the same physical lattice spacing as $\beta=2.4$ 
for the Wilson action.}
\end{figure}

\begin{figure}[!htb]
\begin{minipage}{7.5cm}
\epsfxsize=7cm \epsfysize=7cm
\epsfbox{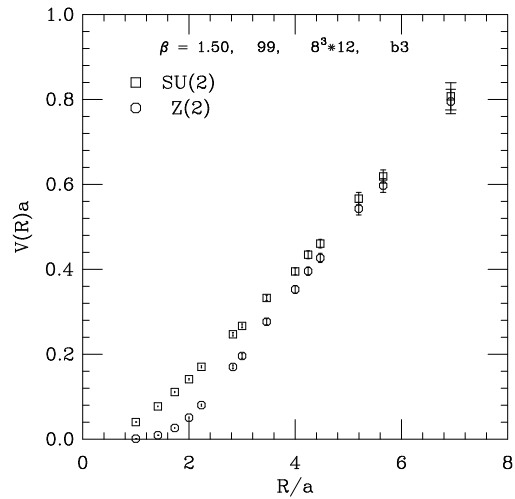}
\end{minipage}
\vspace{-0.7cm}
\caption{\label{fig:pot812_b1.5_b3}The heavy quark potential
 measured after 3 smoothing steps on an ensemble of 
99 $8^3*12$ configurations generated with a fixed point 
action at $\beta=1.5$.}
\end{figure}

Various observables also acquire a physically transparent 
form when expressed in terms of the new variables.
The Wilson loop, in particular, is revealed as essentially 
a vortex counter. It becomes the product of an $SO(3)$ 
part and a $Z(2)$ part that manifestly exhibit the  
flip in sign whenever any of the three types of vortices 
links with the loop. What is surprising, however, is 
that at weak coupling, apart from possible edge (i.e. 
perimeter) contributions, this sign fluctuation appears 
to be all that there is. This suggests 
that the linear piece, i.e. the string tension, if 
nonvanishing, comes {\it entirely} from the expectation of 
the sign flip counting vortices, a rather remarkable 
assertion. We checked this by numerical measurement.
We compared the heavy quark potential extracted from the full 
Wilson loop operator expectation to that extracted from 
the expectation of its sign. 
First using the Wilson 
action, results of  computation of these two quantities, 
labelled $SU(2)$ and $Z(2)$, respectively, are shown 
in figures \ref{fig:pot12_b2.4}, \ref{fig:pot16_b2.5}. 
There is no discernible difference between the two curves.
Note that this coincidence extends down to small Wilson 
loops. This is because we are including {\it all} vortices, 
i.e. also thin ones that can contribute to the 
area law piece of small loops. We next checked this 
by performing the computation 
using a perfect action with a smoothing procedure based on 
the RG \cite{C}. The point is that smoothing removes 
short distance fluctuations while preserving long distance 
physics - in particular, the string tension of the full 
Wilson loop remains unchanged under smoothing. A necessary 
test then of any claim concerning long distance physics 
(here, the claim that replacing the full Wilson loop 
operator by its sign gives the same string tension) is 
that it remain invariant under smoothing.
This is a highly non-trivial test since, in 
general, the smoothed configurations are very 
different from the original 
unsmoothed ones. The results are shown in figures 
\ref{fig:pot812_b1.5_b1} and 
\ref{fig:pot812_b1.5_b3} for one and three smoothing 
steps, respectively. Note that with 
increasing smoothing there is, as it should be, increasing 
deviation at short distances, and in the right direction 
(all thin and generally short vortices are eliminated). 
But the coincidence of the long distance $Z(2)$ and full  
potentials, i.e. the string tension for large loops, 
remains invariant. This is the regime of long thick and 
hybrid vortices at large $\beta$. 
It should be noted that there is a very delicate cancellation 
between positive (even number of vortices) and negative 
(odd number) contributions in the $Z(2)$ expectation that 
conspires to reproduce the asymptotic string tension. 
Conversely, it can be checked that eliminating all (odd 
numbers of) vortices linked with the loop eliminates 
the linear potential. 

Closely related results are reported 
in \cite{G}. We thank J. Greensite for discussions.  

\end{document}